\documentclass[twoside]{article}
\usepackage{graphicx}
\usepackage{fancyhdr}
\usepackage{multicol}
\usepackage{styl-ap}

\def \grs {GRS\,1915+105}
\def \gro {GRO\,J1655$-$40}
\def \gx {GX\,13+1}

\newcommand {\cmmd} {cm$^{-2}$}
\newcommand {\logxi} {$log(\xi)$}

\newcommand {\sigmav} {$\sigma_{\rm v}$}

\def \fetfive {{Fe}\,{\small XXV}}
\def \fetsix {{Fe}\,{\small XXVI}}

\newcommand {\approxgt} {\mathrel{\hbox{\rlap{\lower.55ex \hbox {$\sim$}}
        \kern-.3em \raise.4ex \hbox{$>$}}}}
\newcommand {\approxlt} {\mathrel{\hbox{\rlap{\lower.55ex \hbox {$\sim$}}
        \kern-.3em \raise.4ex \hbox{$<$}}}}

\begin{document}
\title{Disc atmospheres and winds in X-ray binaries}
\author{M. D\'iaz Trigo\work{1} and L. Boirin\work{2}}
\workplace{ESO, Karl-Schwarzschild-Strasse 2, D-85748 Garching bei M\"unchen, Germany
\next
Observatoire Astronomique de Strasbourg, 11 rue de l'Universit\'e, F-67000 Strasbourg, France
}
\mainauthor{mdiaztri@eso.org}
\maketitle

\begin{abstract}%
We review the current status of studies of disc atmospheres and winds in low mass X-ray binaries. We discuss the possible wind launching mechanisms and 
compare the predictions  of the models with the existent observations. We conclude that a combination of thermal and radiative pressure (the latter being relevant
at high luminosities) can explain the current observations of atmospheres and winds in both neutron star and black hole binaries. Moreover, these winds and 
atmospheres could contribute significantly to the broad iron emission line observed in these systems.
\end{abstract}


\begin{multicols}{2}
\section{Introduction}

In the last decade we have witnessed a wealth of discoveries of narrow absorption features in low-mass X-ray binaries (LMXBs). They have been identified with resonant absorption from \fetfive\ and \fetsix\ and other abundant ions and, in a number of systems, are blueshifted, indicating outflowing plasmas. These features were first detected with {\it ASCA} from the microquasars \gro\ (Ueda et al. 1998; Yamaoka et al. 2001) and \grs\ (Kotani et al. 2000; Lee et al. 2002). The launch of the X-ray observatories {\it Chandra}, XMM-Newton and {\it Suzaku}, with ability to  obtain medium to high resolution spectra, opened a new era in studies of plasmas, showing that absorption features are common to black hole (BH) and neutron star (NS) binaries and probably associated to the accretion nature of those systems (e.g. Jimenez-Garate et al. 2002, Boirin et al. 2003, 2004, 2005, D\'iaz Trigo et al. 2006, 2007, 2009, King et al. 2012, Miller et al. 2004, 2006a, 2011, Sidoli et al. 2001, Ueda et al. 2004, 2009, Schulz et al. 2008). 

The overwhelming presence of absorption features in systems that are known to be at high inclination (due to the existence of absorption dips in their light curves) and the modelling of such features with self-consistent photoionised plasma codes led Boirin et al. (2005) and D\'iaz Trigo et al. (2006) to conclude that absorption plasmas are probably ubiquitous to all X-ray binaries, but are only detected in high-inclination systems because the plasma has a flat, equatorial geometry above the disc. 

Studies of photoionised plasmas are important for a number of reasons. Firstly, since such plasmas seem to be ubiquitous in LMXBs, we can learn about accretion processes in these systems by e.g. mapping changes in the disc via changes in the disc atmosphere (or photoionised plasma above the disc). Secondly, since in a fraction of these systems the plasmas are outflowing,
the whole energetic budget of the systems could be significantly altered by the mass expelled in the wind and have an effect in the dynamics of the accretion flow. Thirdly, feedback to the environment could be relevant if the outflows carry a significant amount of kinetic energy and momentum. 

Finally, studies of winds in LMXBs  may provide important information to understand the outflows present in other accretion powered objects. Made up
of a normal star and a collapsed star, LMXBs bridge the gap between Young Stellar Objects and Super Massive Black Holes and may
hold the answers to the driving mechanism of winds. LMXBs are unique in that they show both general relativity effects, when the collapsed star is a 
BH (typically of 5 to 10 solar masses) and effects due to the presence of magnetised stars, when the collapsed star consists of a NS. Thus,
studies of winds in LMXBs allow us to isolate the role of the compact object in driving the winds.

In this work, we review the current observational state and theoretical understanding of disc atmospheres and winds in LMXBs. As a diagnostic tool for this study we focus on the narrow absorption lines, with or without blueshifts. We exclude from our study the winds observed in high mass X-ray binaries, because in these systems the wind from the massive companion star contributes significantly or even dominates the wind component. 

We discuss the scattered component of the wind, its observational evidence and the implications for spin measurements of stellar mass BHs. Finally, we briefly address the potential connection between the presence of compact jets and winds in LMXBs.  

\section{Observational properties and spectral modelling of photoionised plasmas}

To date, narrow absorption features have been discovered in ten NS 
and seven BH LMXBs. 
The lines are consistent with a zero shift for seven systems and show significant outflow velocities consistent with a wind in ten systems. 
However, we note that for two of the systems not showing outflows, the constraints to velocity shifts are weak, $\approxlt$1000~km/s, 
since observations with high resolution spectral gratings are not available (Boirin et al. 2005, Hyodo et al. 2009).

Detailed modelling of the absorption features is often performed for each line individually and followed by a comparison of the line parameters with 
simulations of a photoionised plasma with codes such as XSTAR (Kallman \& Bautista 2001) or CLOUDY (Ferland et al. 1998). A second 
method consists in modelling all the lines and continuum simultaneously with 
self-consistent photoionisation codes such as {\tt warmabs} in XSPEC (Arnaud 1996) or {\tt xabs} in SPEX (Kaastra et al. 1996).
These codes model the absorption due to a photoionised plasma
in the line of sight but taking all relevant ions into account, including those having small cross-sections, which can
contribute significantly to the absorption when combined. The relative
column densities of the ions are coupled through a photoionisation
model.  During the fitting process, {\tt warmabs/xabs} calculate spectra
using stored level populations pre-calculated with XSTAR or CLOUDY for a given ionising continuum.
The main parameters of {\tt
warmabs/xabs} are $N{\rm _H}$, $\xi$, \sigmav, and $v$, representing the
column density of the absorber, the ionisation parameter, the
turbulent velocity broadening, and the average systematic velocity
shift of the absorber (negative values indicate blueshifts). 
The components {\tt warmabs} and {\tt xabs} differ in one important aspect: {\tt xabs} includes self-consistently Compton scattering in the plasma, as opposed to {\tt warmabs}. Electron scattering can affect the whole spectrum significantly when column densities are high. For example, the photoionised plasma present during dips of 4U~1323-62  (N$_H^{xabs} >$\,10$^{23}$ cm$^{-2}$) decreases the flux by more than 50\% below 10~keV and by $\sim$25\% in the 20-50~keV band (Boirin et al. 2005). Therefore, electron scattering should be taken into account in parallel to the use of {\tt warmabs}.
 
The systems showing narrow absorption features are predominantly at high inclination, as first noted for NS (Boirin et al. 2005, D\'iaz Trigo et al. 2006) and recently for BH (Ponti et al. 2012) LMXBs.

For NSs, modelling of the absorption lines with photoionised plasmas yields column densities between 3.5 and 17.2\,$\times$\,10$^{22}$~\cmmd\ and \logxi\,$\sim$~2.5--4.5. The lines are blueshifted in 30$\%$ of the systems, with velocities between $\sim$\,400 and 3000~km/s. 
For BH binaries, the photoionised plasma has column densities between 0.3\,$\times$\,10$^{20}$ and $\sim$\,6\,$\times$\,10$^{23}$\,\cmmd\ and \logxi\,$\sim$~1.8--6. Outflow velocities range between $\sim$\,100 and 1300~km/s and are present in 85$\%$ of the systems. In one case an outflow velocity of 9000-13000 km/s has been claimed, but the significance of the features is low (King et al. 2012).

We note that as observations become sensitive to more lines, the need for more than one ionised plasma is increasing due to the existence of lines with significantly different outflow velocities and the co-existence of lines that require different ionisation equilibria (e.g. Xiang et al. 2009, Kallman et al. 2009). 

For systems showing winds, and considering a purely photoionised plasma, the mass outflow rate 
has been estimated to be of the order of the mass accretion rate (e.g. Ueda et al. 2004, 2009, Ponti et al. 2012), indicating that the winds are important for the dynamics of the system. 

\section{The wind launching mechanism}
\label{launching}

Disc winds from an accretion disc can be launched via thermal, radiative and/or magnetic mechanisms. The launching mechanism may differ significantly for different types of accreting sources and we expect the characteristics of the winds such as geometry, density or outflow velocity to vary depending on the mechanism or mechanisms involved. This will ultimately determine
how much matter can be expelled out of the system and is therefore of utmost importance to assess the dynamical relevance of the winds, both for the systems themselves and for their environment.

UV radiation can be effective in driving a wind because the number of lines is large in this energy range, increasing the probability to transfer energy from UV continuum photons to matter. In X-ray binaries however, X-rays tend to photoionise material and decrease the concentration of ions capable of UV line driving. Therefore, line-driven winds are not expected to be relevant unless shielding from the X-ray irradiation is taking place in some UV-emitting part of the disc (Proga \& Kallman 2002, hereafter PK02).

In contrast, the same X-rays that prevent UV line opacity can heat low density gas to a temperature $\sim$\,10$^7$~K and potentially drive thermal winds.
Assuming the accretion disc is concave and exposed to X-rays from the central source, its upper layers are expected to puff up and expand into an atmosphere, corona or wind (Jimenez-Garate et al. 2002).
The upper boundary of the atmosphere is the Compton temperature corona, which is less dense and hotter than the underlying atmosphere. 
The evaporated photoionised plasma will remain bound to the disc as an atmosphere/corona or be emitted as a thermal wind, depending 
on whether the thermal velocity exceeds the local escape velocity (Begelman et al. 1983, hereafter BMS83; Woods et al. 1996, hereafter W96; PKS02). 
Importantly, the radial extent of the corona is independent of luminosity and determined only by the mass of the compact object and the
Compton temperature (BMS83, W96). W96 determined that a wind would be launched by thermal pressure at radii larger than 0.25\,R$_{IC}$ (where R$_{IC}$ denotes the Compton radius or distance at which the escape velocity equals the isothermal sound speed at the Compton temperature T$_{IC}$). For T$_{IC}$\,$\sim$\,1.3\,$\times$~10$^7$~K, as expected in 
LMXBs,  0.25\,R$_{IC}$ corresponds to a radius of $\sim$1.9\,$\times$\,10$^{10}M$~cm, where $M$ is the mass of the compact object in units of solar masses. However, they also found that even above such radius the wind could be gravity-inhibited if the luminosity were below twice a critical luminosity defined as L$_{cr}\,\sim$\,2.88\,$\times$\,10$^{-2}$\,T$_{IC8}^{-1/2}$\,L$_{Edd}$ (see their eq. 4.4), where L$_{Edd}$ is the Eddington luminosity and T$_{IC8}$ is the Compton temperature in units of 10$^8$~K. Therefore, for  T$_{IC}$\,$\sim$\,1.3 $\times$~10$^7$~K the wind could be gravitationally inhibited for luminosities below 0.16~L$_{Edd}$ (see their Fig. 17). Consequently, for NS systems, at a disc radius of $\sim$2.6\,$\times$\,10$^{10}$~cm an isothermal wind will develop for luminosities above $\sim$\,3$\times$10$^{37}$~erg s$^{-1}$, while such a wind could be inhibited by gravity at lower luminosities. 
For larger radii, a steadily heated free wind could develop already at luminosities below $\sim$\,3$\times$10$^{37}$ erg s$^{-1}$. 

Interestingly, PK02 found that when they included the radiation force, although the line force is dynamically unable to drive a wind, the radiation force due to electrons can be important for very luminous systems: it lowers the effective gravity and subsequently the escape velocity and allows a hot robust disc wind to be already produced at $\sim$\,0.01\,$R_{IC}$, well inside the Compton radius and previous estimates by W96.

Besides the thermal/radiative pressure mechanism described above, in the presence of a strong magnetic toroidal field, magnetic
pressure can give rise to a self-starting wind. 
Calculations of magnetic winds are still scarce and have to make an assumption about the unknown magnetic field configuration. Therefore, it is common practice to 
determine first if a thermal launching mechanism is plausible for a given source and only if this mechanism is ruled out, is the magnetic case considered. 

We note that when magnetic pressure is considered, a wind could be launched well inside the Compton radius and therefore we can use the location of the wind to discriminate between different launching mechanisms. 
\bigskip\\
{\it Comparison of wind models with the current observations of atmospheres and winds in LMXBs}
\bigskip\\
Fig.~\ref{fig:ns} shows the estimated luminosity and location of the photoionised plasma for the NS LMXBs studied with high-resolution gratings or CCD observations. We performed this comparison for NSs rather than for BHs since for the latter the critical luminosity depends on the often uncertain estimates of the BH mass.
For NSs which do not show outflows, we have used the estimations of D\'iaz Trigo et al. (2006) for the luminosity of the system and the location of the plasma (see their Tables 1 and 13), except for Cir~X-1, for which we have used the estimates of Schulz et al. (2008). Uncertainties in the luminosity are derived from the uncertainty in the distance given in Table~13 of D\'iaz Trigo et al. (2006) except for Cir~X-1, for which we use a lower(upper) limit of 6(10.5)~kpc (Schulz et al. 2008, Jonker \& Nelemans 2004). Tight constraints to the existence of an outflow were inferred from grating observations for all the sources shown in Fig.~\ref{fig:ns} except for 4U 1323-62 (Boirin et al. 2005). 

For NSs showing outflows we plot the plasma location and luminosity inferred from grating observations (Schulz \& Brandt 2002, Ueda et al. 2004, Miller et al. 2011). The errors in the luminosity are estimated from the error in the distance and for the radius, a default error of 50\% is adopted whenever the error is not provided in the discovery papers.
The dotted lines have been derived from the lines in Fig.~17 from W96, and set the boundaries for the existence of a wind for NS LMXBs with a T$_{IC}$\,$\sim$ 1.3\,$\times$\,10$^7$~K. The dashed line represents the transition from the corona into a wind when radiation force from electron scattering is included (but note that such limit is applicable only to high luminosities, PK02).    

In short, all the sources that do not show winds (represented with open triangles) are consistent with not showing them if the thermal mechanism is in place, i.e. the location of the photoionised plasma is consistent with the location of the radially bound atmosphere/corona. Moreover, even if the plasma were located at a distance larger than 0.25~R$_{IC}$, their luminosity is below the critical luminosity needed to overcome gravity (W96).
The only exception could be X~1624-490, which has a sufficient large radius and luminosity to launch a wind and for which the average line velocity is consistent with a static atmosphere. Note however that Xiang et al. (2009) find a better fit when two plasmas are used, one of which is outflowing. 

\begin{myfigure}
\centerline{\resizebox{80mm}{!}{\includegraphics{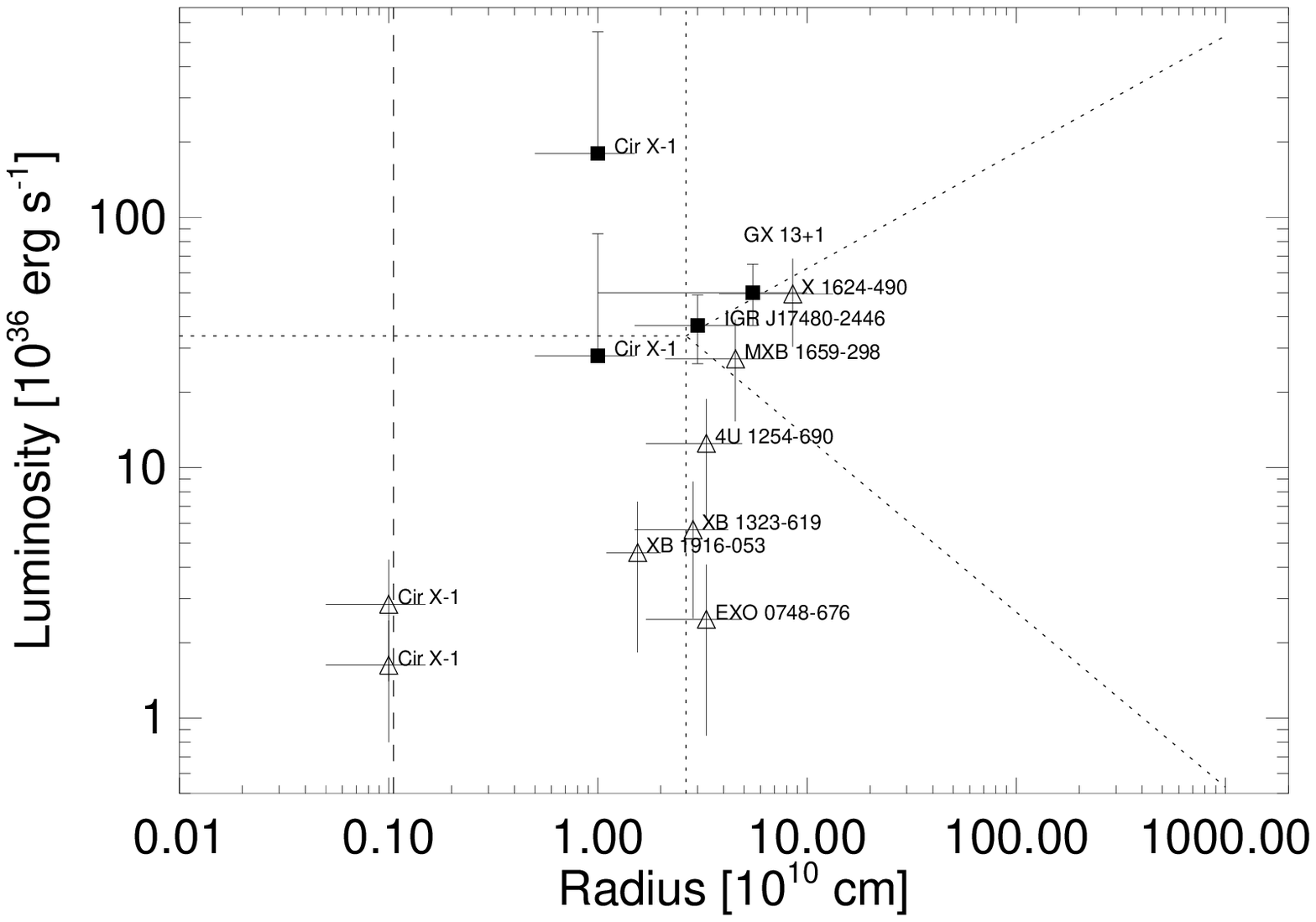}}}
\caption{NS LMXBs for which a photoionised absorber has been detected. Filled and open symbols represent systems with or without an outflow, respectively. The vertical  lines correspond to 0.25~$R_{IC}$ (dotted) and 0.01~$R_{IC}$ (dashed). The horizontal dotted line corresponds to twice the critical luminosity for a NS and marks the boundary between a non-isothermal (bottom) and an isothermal (top) corona. The lower right triangle delimited by the dotted lines shows the region where the wind will be gravity-inhibited.}
\label{fig:ns}
\end{myfigure}

We next look at the sources showing winds, which are represented in Fig.~\ref{fig:ns} with filled squares. It is evident that such sources show luminosities which are significantly larger than for the sources not showing winds, and which are just above twice the critical luminosity derived to launch a wind (W96). In the case of Cir X-1, the launching radius obtained for the wind observed in the high luminosity observations corresponds to $\sim$0.1\,R$_{IC}$. This radius is still plausible for launching winds once the radiation pressure due to electron scattering is considered (see PK02), especially at this high luminosity. In summary, Fig.~\ref{fig:ns} indicates that a thermal mechanism explains satisfactorily the presence or absence of winds in NS LMXBs. Further support to this conclusion is given by the fact that winds and warm absorbers are preferentially detected in high-inclination sources (all the sources shown in Fig.~\ref{fig:ns} show dips in their light curves, indicating inclinations larger than $\sim$\,65$^\circ$). This is consistent with the geometry predicted for a thermal wind, which will be observed close to the equatorial plane, since at polar
angles, $\approxlt$\,45\,$^{\circ}$, the low density and the high ionisation of the gas prevent its detection.

A caveat to this interpretation is that the Compton temperature for the different sources may differ from the one considered above. To evaluate the uncertainty introduced by the assumption of T$_{IC}$\,$\sim$\,1.3\,$\times$\,10$^7$~K, we calculated T$_{IC}$ for three of the sources for which models fitted to broadband X-ray spectra were found in the literature.
We obtained values of $\sim$ 0.3, 0.4 and 1$\times$10$^7$ K for 4U~1254-69, X~1624-49 and GX~13+1, respectively. These values imply that R$_{IC}$ could be $\sim$\,3--5\,$\times$10$^{11}$~cm for the first two sources placing the plasma well inside the calculated radial extent of the corona.

We note that uncertainties in the derivation of the location of the plasma are mainly driven by uncertainties in its density. A second source of uncertainty is the bolometric luminosity of the source, which is inferred from the X-ray luminosity and may be severely underestimated when there is a high opacity in the line of sight, which happens precisely for sources observed at high inclination and with a strong wind. 

We conclude that the presence or absence of winds in current observations of NS LMXBs can be explained by thermal/radiative pressure in the disc, when all the uncertainties are taken into account.
For BH LMXBs, the uncertainty in the mass of the BH increases the uncertainties in the calculation of the Compton radius and the critical luminosity and make more difficult to discriminate between different launching mechanisms. To date there is only one claim of a magnetic driven wind in a BH LMXB (Miller et al. 2006a), which has been controversial (Netzer et al. 2006). GRO\,J1655--40 was extensively monitored by X-ray observatories during its 2005 outburst. XMM-Newton observations showed a highly ionised wind consistent with a thermal mechanism (D\'iaz Trigo et al. 2007). However, the wind had strongly evolved at the time of the $Chandra$ observation performed five days after the last XMM-Newton observation. More than 70 lines were present and revealed a very dense wind. The most exhaustive study of this spectrum was performed by Kallman et al. (2009), who derived a radius of 7\,$\times$\,10$^9$~cm for the location of the plasma and concluded that the wind could not be thermally launched.  However, given the uncertainty in the
the density of the plasma (10$^{13}$--10$^{15}$ cm$^{-3}$) and the bolometric luminosity (since obscuration of the continuum due to Compton scattering in the wind was not considered), we conclude that it is too early to discard a thermal launching mechanism for this wind. Different Compton temperatures, obtained by fitting plausible continuum models to broadband spectra, should also be considered when calculating the 
allowed radii. In summary, even if several mechanisms could be in place at the same time, we find that magnetic pressure is not requested to explain the existent observations of winds.

\section{Scattering in the wind}

The simultaneous presence of narrow absorption features, indicating the presence of a warm absorber, and of a broad iron line has been often observed in high inclination, dipping, LMXBs (e.g. Sidoli et al. 2001, Parmar et al. 2002, Boirin et al. 2005, Kubota et al. 2007, D\'iaz Trigo et al. 2009, 2012). Due to the symmetry and relatively modest size of the broad iron line, its origin has been commonly attributed to the combination of line blending, Doppler broadening and Compton scattering in an accretion disc corona or hot atmosphere. Recently, XMM-Newton observations of the NS LMXB \gx\ have shown a correlation of the variations of the broad iron line and the state (column density and degree of ionisation) of the warm absorber, indicating that absorption and emission take place most likely in the same plasma (D\'iaz Trigo et al. 2012). 

These observations are in agreement with models of photoionised winds by Sim et al. (2010a, 2010b). In these models absorption
features are imprinted in the spectra when we look through the wind and the radiation scattered in the outflow produces the broad iron line emission and
other distinct features such as a Compton hump, which will be more or less visible depending on their contribution with respect to the incident radiation. 
At low, $\approxlt$45\,$^\circ$, inclinations the wind does not obscure the X-ray source and consequently, absorption features are not observed. However, a significant component of scattered/reprocessed radiation in the outflow is present in addition to the direct emission and responsible for the broad iron line emission. The emission is predominantly formed by \fetfive\ and \fetsix, since the part of the wind seen at low inclinations is the highly ionised surface of the outflow.

Given that warm absorbers and/or winds seem to be ubiquitous to all LMXBs (Boirin et al. 2005, D\'iaz Trigo et al. 2006, Ponti et al. 2012), and that the scattered component of the wind should be visible at all inclinations (e.g. Sim et al. 2010a), it follows that broad line emission should be observable in a majority of LMXBs. This is consistent with systematic analyses of broad iron emission lines in NS LMXBs (see Ng et al. 2010 and references therein), where lines are found in 50--85$\%$ of the sources and are highly ionised.

A systematic study of broad iron lines in BH LMXBs is challenging due to the transient nature of the sources. However, given the similarity of winds in NS and BH LMXBs, it is not unexpected that at least a fraction of the broad iron lines observed in BH LMXBs are produced at the wind and not by reflection at the inner disc, as it is often assumed. Therefore, measurements of the BH spin based on the breadth of the iron emission line will have an uncertainty associated to the contribution of the wind to the line, which may be comparable or even larger than the component arising from reflection at the inner disc. In conclusion, 
our ability to probe the physical conditions at the inner disc depends on our understanding of accretion disc physics, including atmospheres and winds.

\section{A wind-jet connection?}

Disc winds have been observed in the high/soft, thermal-dominated, state of BH transients, when
the jet emission is absent. In contrast, in observations of the low/hard state of the same transients,
with typical jet emission, winds were excluded, indicating that there may be an anticorrelation
between winds and jets. In particular, blueshifted absorption lines were observed in
the soft state spectra of \gro\ (Miller et al. 2006a, D\'iaz Trigo et al. 2007),
\grs\ (Ueda et al. 2009), GX339--4 (Miller et al. 2004), 4U1630--472 (Kubota et
al. 2007) and H1743--322 (Miller et al. 2006b). In contrast, observations of the hard state
of \gro\ (Takahashi et al. 2008) and H1743--322 (Miller et al. 2006b) excluded the presence of such lines
down to equivalent widths of 20 and 3 eV, respectively. The absence of strong winds in the hard state 
of BH LMXBs has been recently confirmed by a systematic search of absorption lines in existent X-ray
observations (Ponti et al. 2012). 
 
Based on \grs\ observations, Neilsen \& Lee (2009) proposed that a possible explanation
for the observed anti-correlation between jets and winds is that the wind observed during
the soft state carries enough mass away from the
disc to halt the flow of matter into the radio jet.
However, there is one detection of a weak wind in
the ``hardÓ (or ``CÓ) state of \grs\ (Lee et al. 2002). Therefore, it could also be possible
that a high amount of matter is carried away in both soft and hard states, but due to the higher
ionisation in the hard state the ions become fully stripped and the wind becomes transparent and
thus ``invisible". 

Although significant changes in the ionisation and column density of the wind with X-ray luminosity
have been already observed during the soft states of BH outbursts
(e.g. D\'iaz Trigo et al. 2007, Kubota et al. 2007), a systematic study of the evolution of the Compton
temperature and the wind throughout the whole outburst has to be done yet and is now of utmost importance 
to determine the reason for the absence of wind detections in the hard state. 

\section{Discussion and conclusions}

The overwhelming presence of hot atmospheres and winds in high inclination LMXBs shows that they are most likely situated in a flat geometry around the disc and are therefore ubiquitous in LMXBs. Modelling of the absorption features with photoionised plasmas is consistent with a scenario in which the disc is heated by irradiation from the central object and an atmosphere and corona are formed, which can be observed as an outflowing wind under certain conditions of distance to the central object and luminosity. 

The reprocessed/scattered component of the wind has been only scarcely studied. The high plasma column densities observed in some objects indicate that opacity in the wind could be relevant and scattering in the plasma a major source of broad line emission (Sim et al. 2010a and references therein). In this respect, it is now of utmost importance to include the reprocessed
component in wind models to compare them with the observations and be able to quantify their significance for BH spin measurements based on the broad iron emission line. 

Finally,  the appearance or disappearance of winds could be triggered by changes in the structure of the accretion flow. Sensitive simultaneous observations in X-ray and radio wavelengths of BH binaries throughout their outbursts together with an estimation of the expected winds at each state could soon shed some light on this topic.

Future observations of winds with high resolution calorimeters, as the one onboard {\it Astro--H}, will allow us to determine with unprecedented accuracy the sites of wind production, thus further constraining the dynamics of accretion flows.

\end{multicols}
\end{document}